\documentclass[aps,prb,twocolumn,showpacs,groupedaddress]{revtex4}

\usepackage{graphicx}

\begin{document}

\preprint{}

\title{Reversible room-temperature ferromagnetism in Nb-doped SrTiO$_{3}$ single crystals}

\author{Z. Q. Liu$^{1,2}$}

\author{W. M. L\"{u}$^{1,3}$}

\author{S. L. Lim$^{2}$}

\author{X. P. Qiu$^{3}$}

\author{N. N. Bao$^{1}$}

\author{M. Motapothula$^{1,4}$}

\author{J. B. Yi$^{5}$}

\author{M. Yang$^{2}$}

\author{S. Dhar$^{1,3}$}

\author{T. Venkatesan$^{1,2,3}$}

\author{Ariando$^{1,2}$}

\altaffiliation[Email: ]{ariando@nus.edu.sg}

\affiliation{$^1$NUSNNI-Nanocore, National University of Singapore, 117411 Singapore}

\affiliation{$^2$Department of Physics, National University of Singapore, 117542 Singapore}

\affiliation{$^3$Department of Electrical and Computer Engineering, National University of Singapore, 117576 Singapore}

\affiliation{$^4$Center for Ion Beam Applications, National University of Singapore, 117542 Singapore}

\affiliation{$^5$Department of Material Science and Engineering,
National University of Singapore, 117576 Singapore}

\date{\today}

\begin{abstract}
The search for oxide-based room-temperature ferromagnetism has been
one of the holy grails in condensed matter physics. Room-temperature
ferromagnetism observed in Nb-doped SrTiO$_{3}$ single crystals is
reported in this Rapid Communication. The ferromagnetism can be eliminated by air
annealing (making the samples predominantly diamagnetic) and can be recovered by subsequent vacuum annealing.
The temperature dependence of magnetic moment resembles the temperature dependence of carrier density, indicating that the magnetism is closely related to the free carriers. Our results suggest that the ferromagnetism is induced by oxygen vacancies. In addition, hysteretic magnetoresistance was observed for magnetic field parallel to current, indicating that the magnetic moments are in the plane of the samples. The x-ray photoemission spectroscopy, the static time-of-flight and the dynamic secondary ion mass spectroscopy and proton induced x-ray emission measurements were performed to examine magnetic impurities, showing that the observed ferromagnetism is unlikely due to any magnetic contaminant.
\end{abstract}

\pacs{75.47.Lx, 75.50.Pp, 75.70.Rf}


\maketitle


The potential for discovering new magnetic interactions with
possible applications in spintronic devices has been the main driver
for the search of oxide-based room-temperature ferromagnetism (RTFM)
[1,2]. Since the theoretical prediction of  RTFM  in Mn-doped ZnO
[3] and the experimental observation of  RTFM in Co-doped TiO$_{2}$
[4], dilute magnetic semiconductors (DMS) has attracted significant
attention from the community of oxide electronics. Typically, DMS
are fabricated by introducing magnetic ions into wide bandgap
semiconductors including ZnO , TiO$_{2}$, SnO$_{2}$ [5] and
In$_{2}$O$_{3}$ [6]. In addition, since the finding of unexpected
ferromagnetism in insulating HfO$_{2}$ thin films [7], RTFM was also observed in pristine
TiO$_{2}$ [8,9], In$_{2}$O$_{3}$ [9], ZnO [10,11] and SnO$_{2}$ [12]
thin films without magnetic dopings, which is generally attributed
to oxygen vacancies or other ionic defects. SrTiO$_{3}$ (STO) is
important for oxide electronics due to its chemical and
thermal stabilities, as well as the lattice match to a large number
of functional perovskite materials. Pristine STO is a typical
nonpolar band insulator with an indirect band gap of 3.25 eV and a
direct band gap of 3.75 eV [13]. However, the slight doping of Nb
can shift the Fermi level of STO up or even into the bottom of its
conduction band, thus giving an \emph{n}-type semiconducting or metallic
phase [14]. Nb-doped SrTiO$_{3}$ (NSTO) itself is a very interesting
system in that it is a low-temperature two-band superconductor [15],
with strong interactions among electrons, plasmons, phonons and
polarons [16,17]. More importantly, it is one of the most used
substrates for oxide film deposition and oxide electronic device
applications [18].

A pristine STO single crystal is an ideal diamagnet due to the
absence of unpaired electrons. The Nb doping replaces some of Ti
atoms and the resultant Ti$^{3+}$ ions with unpaired electrons can
generate a basic paramagnetic response to an external magnetic
field. Here we report RTFM observed in NSTO single-crystal substrates, which is found to be induced by oxygen vacancies
and possibly mediated by free electrons from Nb doping. We examined
the transport and magnetic properties of 5 mm$\times$5 mm$\times$0.5
mm NSTO single crystals with different dopings, \emph{i.e.}, 0.05
wt\% 0.1 wt\%, 0.5 wt\% (CrysTec GmbH, Germany), 0.7
wt\% (Hefei Kejing Material Technology Co., Ltd.,
China) and 1 wt\% (MTI, USA). The transport and
magnetic properties were measured in a Quantum Design PPMS and a
Quantum Design SQUID-VSM tool, respectively.


\begin{figure}
\includegraphics[width=2.5in]{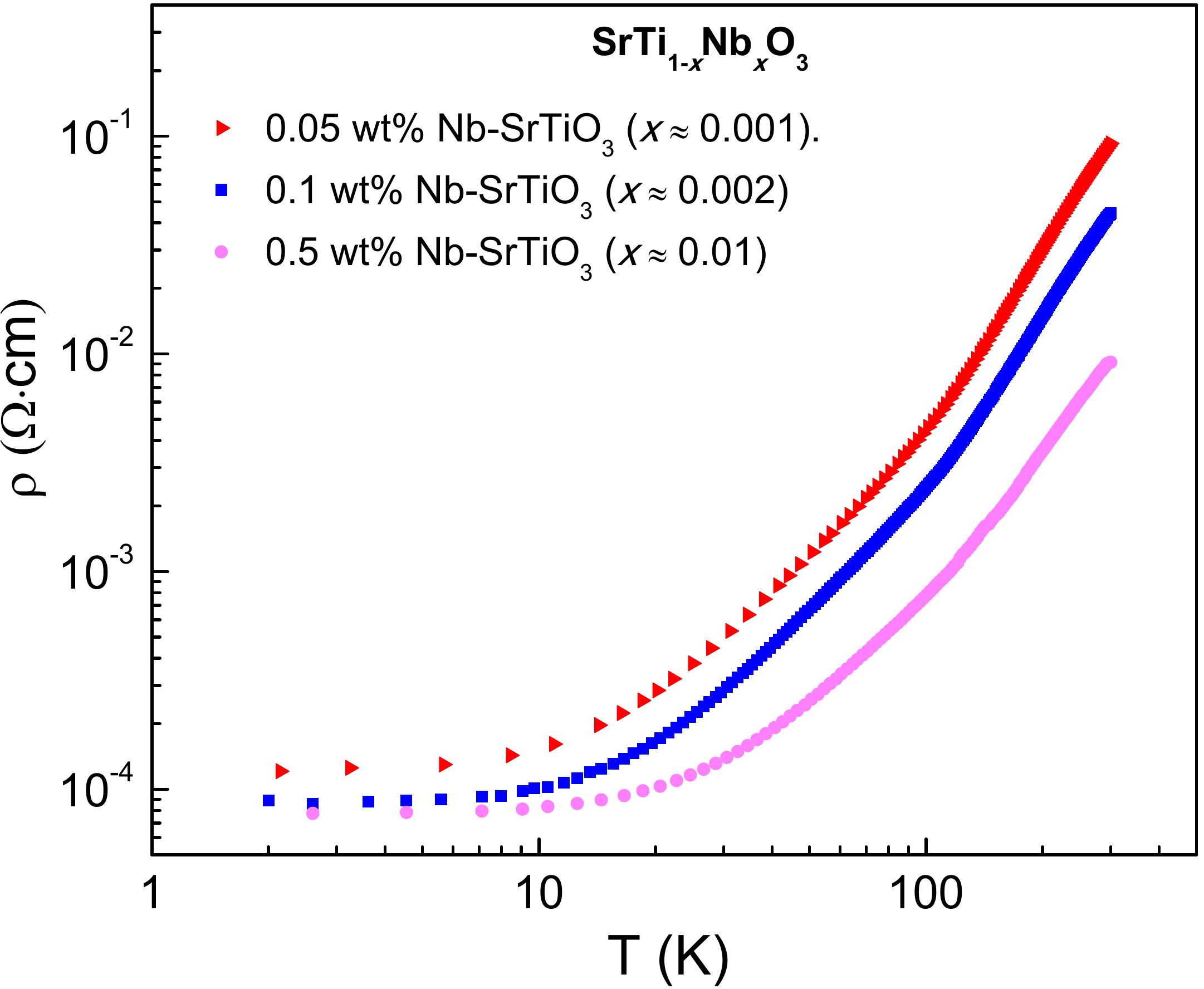}
\caption{\label{fig1} (Color online) Temperature dependence of
resistivity of Nb-doped SrTiO$_{3}$ (NSTO) single crystals with
different dopings.}
\end{figure}

The temperature dependences of resistivity from 300 K to 2 K for
different NSTO single crystals are shown on a logarithmic scale in
Fig. 1. All of them indicate the typical metallic behavior and the
resistivity of NSTO decreases with the doping level over the whole
temperature range. The moment versus temperature (\emph{M-T})
measurements were performed for NSTO single crystals by a 1000 Oe
magnetic field applied parallel to the surface of samples. The
\emph{M-T} curves of NSTO single crystals are shown in Fig. 2. For
0.05 wt\% and 0.1 wt\% dopings, negative magnetic moment over the
whole temperature range in Fig. 2(a) indicates that the diamagnetism
from STO matrix is dominant. The magnetic moment versus magnetic
field (\emph{M-H}) curves measured up to 2000 Oe show linear field
dependence of moment even at low temperatures down to 2 K as seen in
the inset of Fig. 2a. That demonstrates only diamagnetism and
paramagnetism coexist in 0.05 wt\% and 0.1 wt\% NSTO single
crystals, where paramagnetism leads to the slight increase of
magnetic moment at low temperatures.

The \emph{M-T} curve of a 0.5 wt\% NSTO single crystal is shown in
Fig. 2(b), which is evidently distinct from the \emph{M-T} curves
for single crystals with lighter doping. Unexpectedly, the magnetic
moment peaks at $\sim$60 K and ferromagnetic hysteresis loops can be
seen from 2 K to 300 K as shown in Fig. 2(c). The average coercivity
field is between 300 and 400 Oe. The saturation ferromagnetic moment at 2 K
is $\sim$5$\times$10$^{-6}$ emu and the corresponding magnetization
for the whole 5 mm$\times$5 mm$\times$0.5 mm single crystal is
4$\times$10$^{-4}$ emu/cm$^{3}$, which is equivalent to
$\sim$5$\times$10$^{-3}$ Oe. The average
ferromagnetic moment for each Nb atom (or roughly each free
electron) of NSTO is $\sim$2.65$\times$10$^{-4}$
$\mu_{B}$, which is two orders of magnitude smaller than the ferromagnetic
moment of free-electron gas, 0.07 $\mu_{B}$/electron reported by
Young \emph{et al.} [19].

\begin{figure}
\includegraphics[width=3.4in]{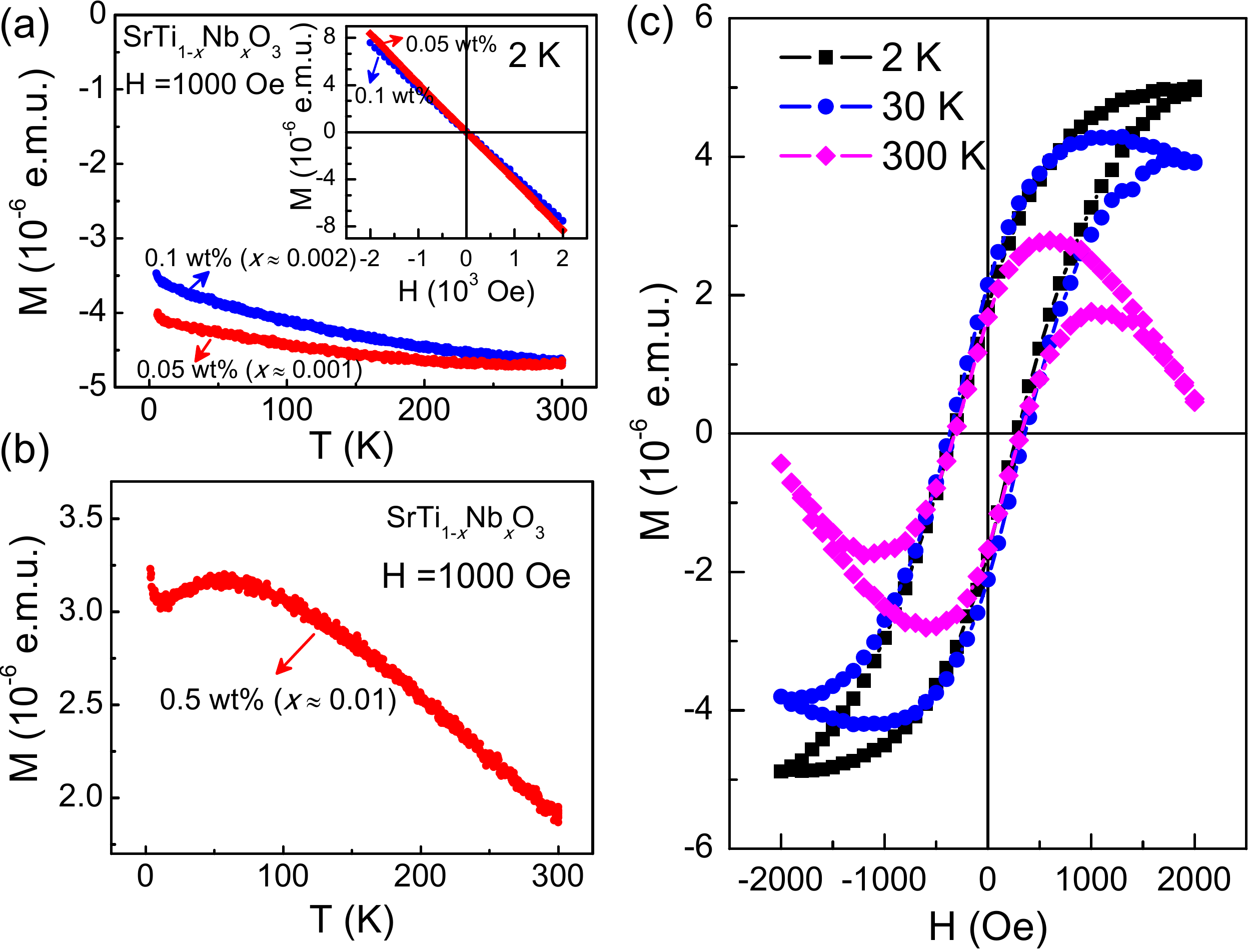}
\caption{\label{fig2} (Color online) Magnetic moment of 5
mm$\times$5 mm$\times$0.5 mm NSTO. (a) Magnetic moment versus
temperature (\emph{M-T}) curves of 0.05 wt\% and 0.1 wt\% NSTO.
(Inset) \emph{M-H} curves at 2 K. (b) \emph{M-T} curve of 0.5 wt\%
NSTO. (c) Magnetic moment versus magnetic field (\emph{M-H}) curves
of 0.5 wt\% NSTO at 2 K, 30 K and 300 K.}
\end{figure}

NSTO single crystals from different batches and different vendors were
investigated. It was found that all of 0.5 wt\% samples (CrysTec GmbH, Germany) show ferromagnetic hysteresis loop
with the same order of magnitude of the magnetic moment at 2 K as
well as 300 K [20]. Moreover,  the 0.7 wt\% NSTO single crystal
(from Hefei Kejing Material Technology Co., Ltd., China) and 1 wt\%
NSTO (from MTI, USA) also show similar ferromagnetic hysteresis
loops both at 2 K and 300 K [20]. Assuming that the ferromagnetism
originates from some ferromagnetic artifacts, like Fe, which can
maximally supply 2.2 $\mu_{B}$/atom to ferromagnetism, the
corresponding minimum density of Fe impurity in 0.5 wt\% NSTO is
$\sim$1.96$\times$10$^{16}$ atoms/cm$^{3}$.  It is within the
typical detection limit of static time-of-flight secondary ion mass
spectroscopy (SIMS) [21]. SIMS analysis of possible Fe, Co, Ni, Cr,
Mn and Cu elements were performed down to more than 400 nm below the
surface at different regions of the sample but no trace of them was
shown [20]. In addition, x-ray photoemission spectroscopy (XPS)
examination with Ar ion milling was carried out down to more than
100 nm below the surface. The wavelength regions of the
characteristic photoemission peaks of possible Fe, Co and Ni
elements were carefully examined but also no any signature of them was
seen [20].

Later it was found the ferromagnetism can greatly decrease or even
disappear after annealing single crystals in air. The \emph{M-H}
hysteresis at 2 K and 300 K completely disappear after annealing at 600 $^{\circ}$C for 2 h, as shown in
Fig. 3(a). Subsequently, the sample was vacuum-annealed
at 950 $^{\circ}$C in $\sim$10$^{-7}$ Torr vacuum for 1 h and the ferromagnetism was recovered [Fig. 3(b)]. Upon fitting and subtraction of the diamagnetic
signal, the pure ferromagnetic loop can be extracted from the
\emph{M-H} loop up to 1 T [20]. The final saturation ferromagnetic
moment recovered from vacuum annealing is $\sim$8.5$\times$10$^{-6}$
emu, which is of the same order of magnitude with the ferromagnetic
signal in the original case. These results indicate that the
ferromagnetism is closely related to oxygen vacancies in NSTO
single crystals. Moreover, it was found that the ferromagnetic moment in an as-received 0.5 wt\% NSTO single crystal significantly decreases even after annealing in air at 250 $^{\circ}$C for 30 mins [20]. As the diffusion coefficient \emph{D} of oxygen ion in STO below 450 $^{\circ}$C is less than 10$^{-16}$ cm$^{2}$/s [22], the diffusion length \emph{l} = $\sqrt{\emph{Dt}}$ is therefore less than 40 nm for 30 mins. This indicates that the ferromagnetic moments mostly exist in the surface region of NSTO single crystals.

\begin{figure}
\includegraphics[width=3.4in]{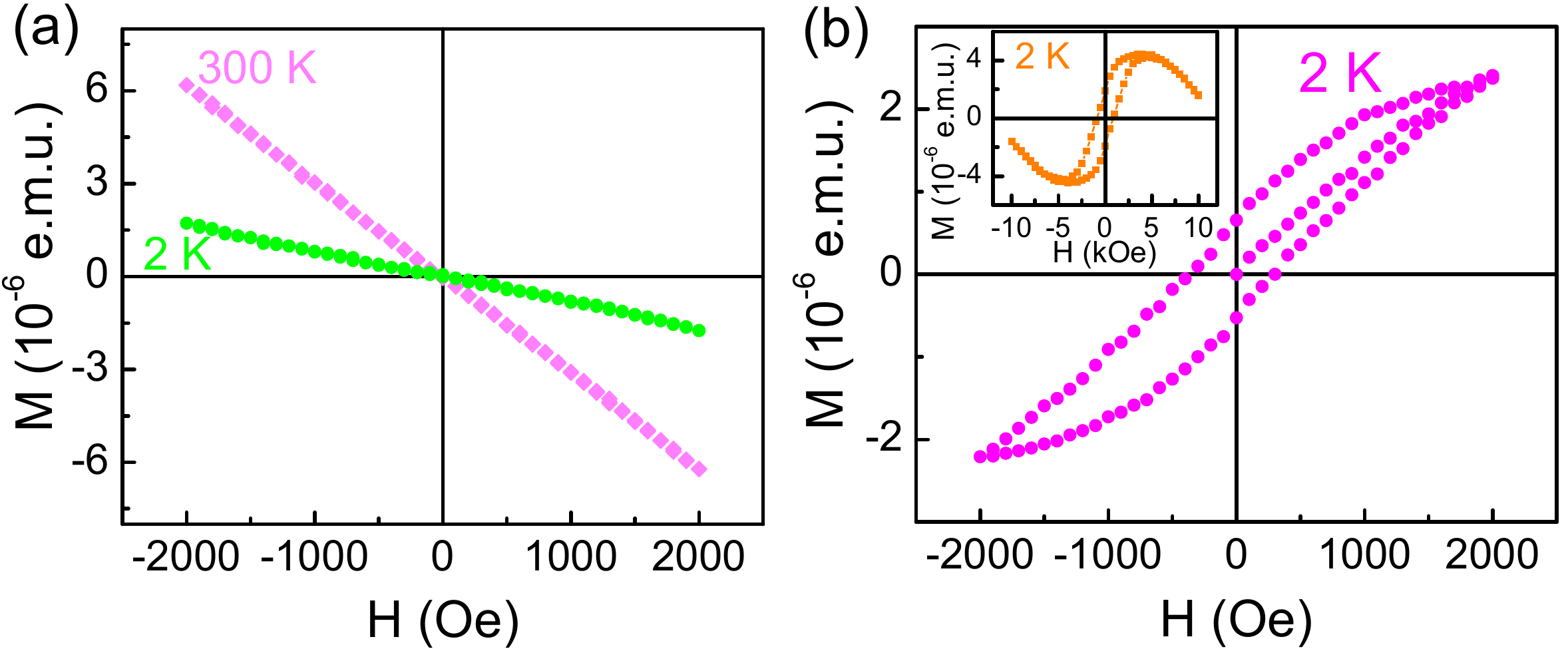}
\caption{\label{fig3}(Color online) Magnetic moment of 0.5 wt\% NSTO for
continuous annealing procedures. (a) \emph{M-H} curves at 2 K and 300 K after annealing
at 600 $^{\circ}$C in air for 2 h. (b) \emph{M-H} curves at 2 K upon the subsequent vacuum annealing at
950 $^{\circ}$C in $\sim$10$^{-7}$ Torr vacuum for 1 h.
(Inset) \emph{M-H} loop at 2 K measured up to 1 T. }
\end{figure}

To further examine possible impurities of Cr, Mn, Fe, Co, Ni and Cu in vacuum-annealed samples, we performed
dynamic SIMS measurements using a Cameca IMS-6f magnetic sector spectrometer
with a sensitivity of ppb, which is three orders of magnitude higher than the
sensitivity of the typical static time-of-flight SIMS. During the analysis, a Cs$^{+}$ primary ion beam of 10 keV was rastered over an area of 250$\times$250 $\mu$m$^{2}$, with the samples biased at a
voltage of +5 kV. Positive secondary ions were acquired from a central area
of approximately 40 $\mu$m in diameter. Depth profiling spectra were first acquired
over a thickness of approximately 500 nm, within which the matrix elements
were observed to be uniform as shown for a vacuum-annealed sample in Fig. 4(a).
At the specified depth, mass spectra were then collected over the mass range
of interest (45-70 a.m.u.), which confirmed the absence of Cr, Mn, Fe, Co
and Ni impurities [Fig. (4b)]. Due to the mass interference of $^{63}$Cu
with $^{47}$Ti$^{16}$O, high resolution mass spectra
were acquired at the mass range of $^{47}$Ti$^{16}$O (with mass resolution higher
than 5000). No signal of $^{63}$Cu was detected as well as indicated by Fig. 4(c). Since the SIMS experiment only probes the surface of the samples (around 500 nm in depth), we performed proton induced x-ray emission experiments (PIXE) where the 2.1 MeV proton beam is able to probe a depth of 60 $\mu$m. The dominant impurities were found to be Fe and Zn at 2.1 and 0.4 at\%, respectively, for the oxygen annealed samples (diamagnetic), while for vacuum annealed samples (ferromagnetic) the concentration were 1.5 and 1.3 at\%, respectively. Clearly, these impurities are highly mobile during the various annealing processing steps, but they do not show any correlation with respect to the observed magnetism. This is consistent with the experimental where 2 at\% magnetic elements (Fe, Cr, Co and Mn) were intentionally added to epitaxial NSTO films but no ferromagnetism was seen [23].

\begin{figure}
\includegraphics[width=3.4in]{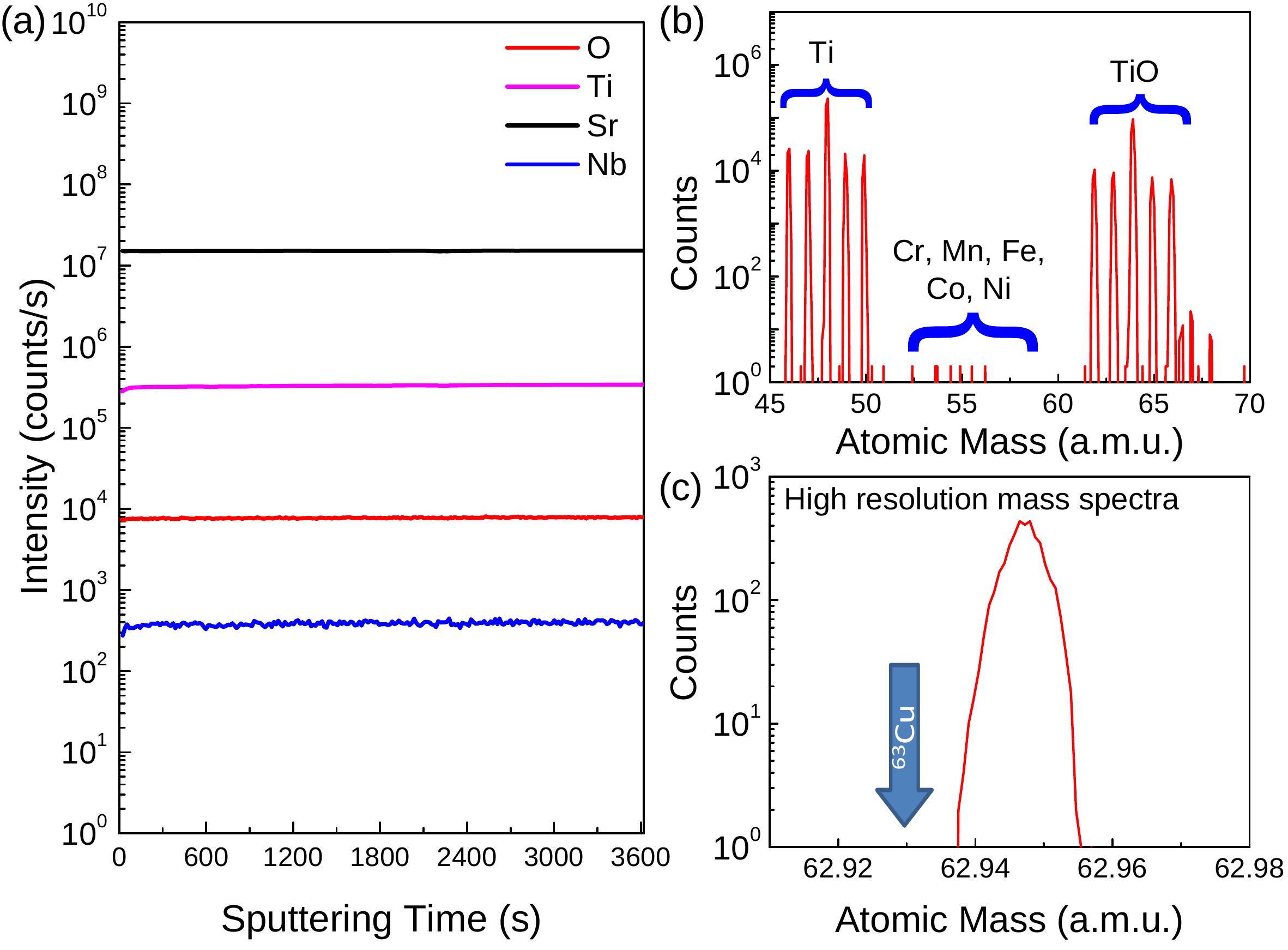}
\caption{\label{fig4} (Color online) Dynamic SIMS (a) Depth profiling spectra of
a vacuum-annealed 0.5 wt\% NSTO single crystal. (b) Mass spectra over the
mass range of 45-70 a.m.u. (c) High resolution mass spectra at the mass range of $^{47}$Ti$^{16}$O.}
\end{figure}

The magnetoresistance (MR) of 0.5 wt\% NSTO single crystals were studied. Transverse MR is positive and quadratically depends on magnetic field [20], which suggests that the orbital scattering dominates due to the Lorentz force. No hysteresis in transverse MR curves was observed. Nevertheless, as the magnetic field is applied parallel to the current, negative MR shows up. More importantly, there is a hysteresis loop in the MR curve (Fig. 5), typical of ferromagnetic materials [24]. Nevertheless, the MR loop here exhibits double peaks in a single round field scan from -9 T to 9 T or 9 T to -9 T, which is different from that of conventional ferromagnetic materials [24]. The double peaks could be that STO is a high mobility system and accordingly positive MR from orbital scattering could dominate at low field if the current is not well-confined to be exactly parallel to magnetic field, which are commonly observed in STO-based electron systems [25-28]. Moreover, the saturation field in the ferromagnetic hysteresis loop is consistent with the closure field in the MR data. The hysteresis reveals that the ferromagnetism observed in NSTO single crystals is intrinsic.

\begin{figure}
\includegraphics[width=2.5in]{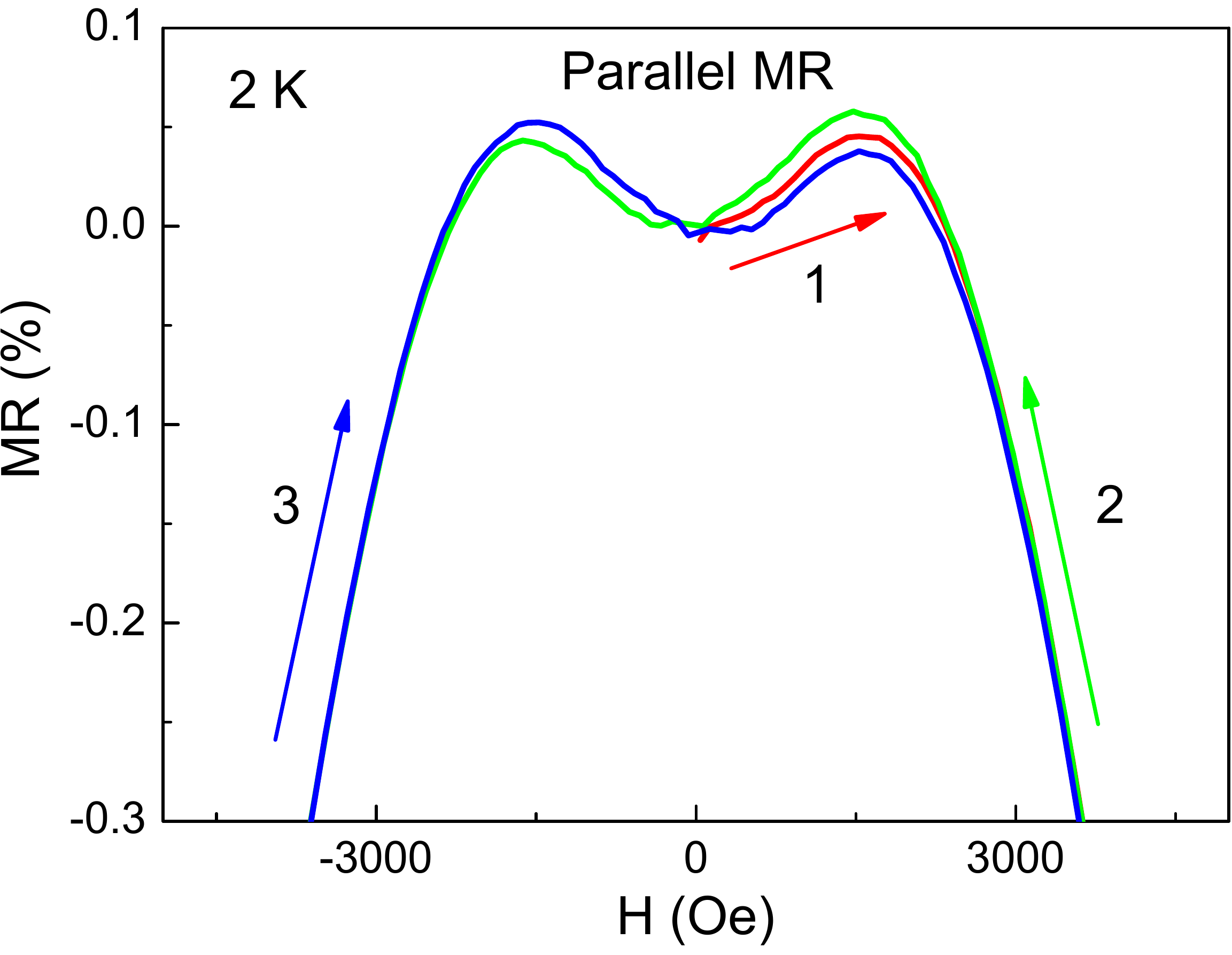}
\caption{\label{fig5} (Color online) Parallel MR of a 0.5 wt\% NSTO single crystal at 2 K. The numbers 1, 2 and 3 indicate the measurement sequence.}
\end{figure}

The temperature dependence of carrier density (\emph{n-T}) for as-received NSTO single
crystals were measured via Van der Pauw Hall geometry with two
voltage and two current electrodes placed at the four corners of 5 mm $\times$ 5 mm
square single crystals. No anomalous Hall effect was obtained, which is likely because the ferromagnetic moment mostly existing in the surface region is too small for an entire single crystal to generate the anomalous Hall effect. Intriguingly, it was found that the carrier density of NSTO
first increases with lowering temperature, and then peaks at a certain temperature
as shown in Fig. 6(a). Meanwhile, the structural phase transition of bulk STO at $\sim$105 K
can be clearly seen from the linear fittings of the temperature dependence of mobility
in Fig. 6(b), which was also observed in oxygen-deficient STO single crystals [29].
The peak temperature depends on the doping level. For 0.5 wt\% NSTO,
the peak is at $\sim$60 K and also the \emph{n-T} curve is similar to the \emph{M-T} curve
in Fig. 2(b). After annealing in air, the \emph{n-T} curve does not substantially change because the
carrier density generated from oxygen vacancies of 1 h vacuum annealing in STO is
of the order of 10$^{18}$ cm$^{-3}$ carriers [29], which is two orders of magnitude smaller than
the carrier density from Nb doping for 0.5 wt\% NSTO. However the original peak
at $\sim$60 K in the \emph{M-T} curve disappears, accompanied by the disappearing of the
ferromagnetic hysteresis loop. No similarity between the \emph{n-T} and \emph{M-T}
curves for 0.05 wt\% and 0.1 wt\% single crystals was observed [inset of Fig. 6(a)]. The similarity between the \emph{n-T} and \emph{M-T} curves in 0.5 wt\% NSTO
indicates that the ferromagnetism is sensitive to free carriers.

\begin{figure}
\includegraphics[width=3.4in]{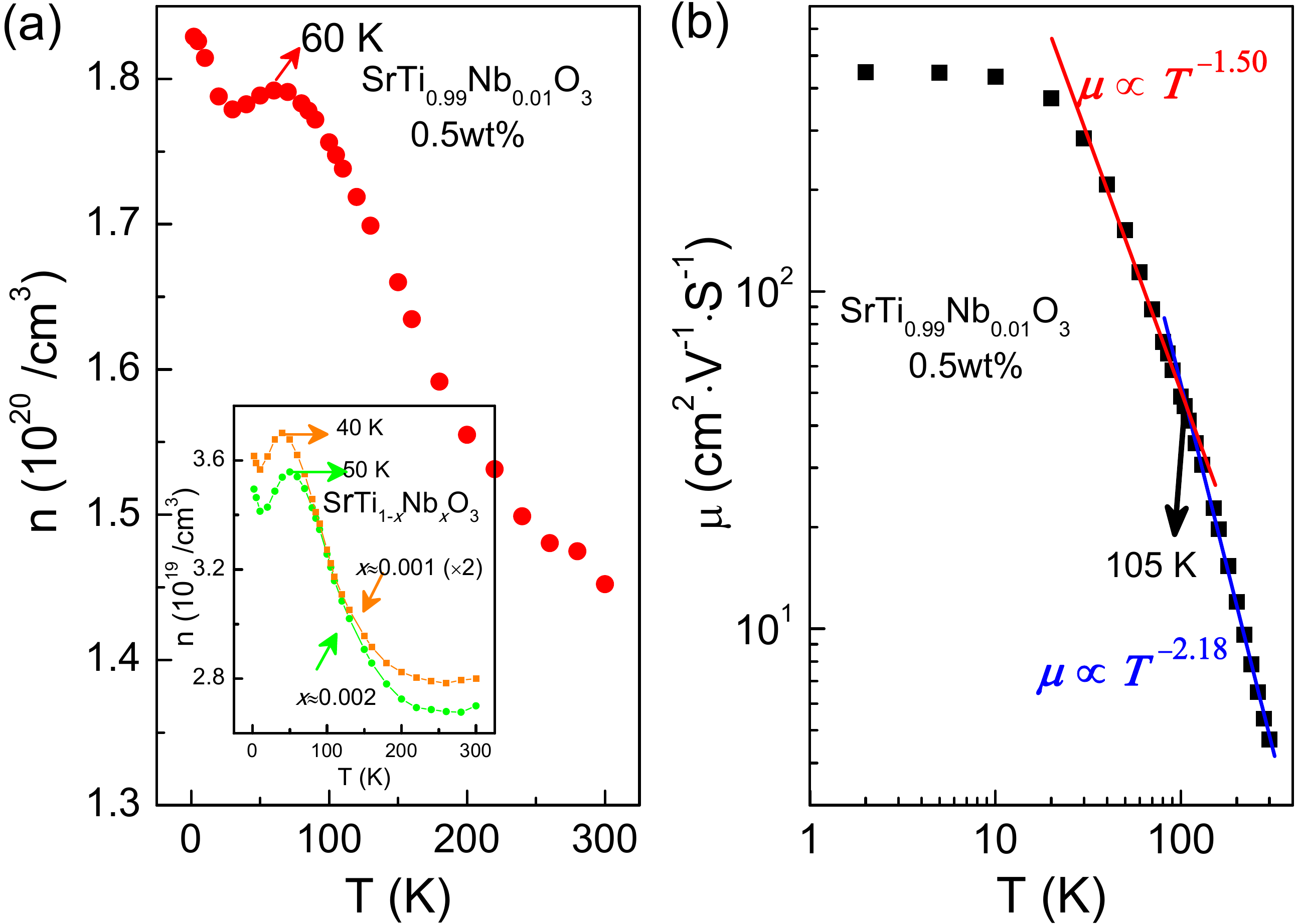}
\caption{\label{fig6} (Color online) (a) Temperature dependence of
carrier density (\emph{n-T}) of NSTO single crystals. (Inset) \emph{n-T}
curves of 0.05 wt\% and 0.1 wt\% NSTO. The carrier density of 0.05 wt\% NSTO is magnified by a factor of two. (b) Temperature dependence of
mobility ($\mu$) of the 0.5 wt\% NSTO single crystal.}
\end{figure}

For as-received 0.05 wt\% and 0.1 wt\% NSTO single crystals, no
ferromagnetism was observed. The reason may be either the carrier
density is not large enough or the initial oxygen vacancies inside
are less. Since Nb$^{5+}$ ions can attract O$^{2-}$ more and therefore deform
the crystal lattice, therefore a larger doping could generate more
defects including oxygen vacancies in the lattice. Indeed, ferromagnetism was also seen in vacuum-annealed 0.05 wt\% and 0.1 wt\% NSTO single crystals [20]. Moreover, it was found that the saturation ferromagnetic moment of vacuum-annealed NSTO single crystals almost linearly scales with the doping level [20]. This further emphasizes the important role of carrier density in the ferromagnetism of NSTO.

Although both oxygen vacancies and Nb atoms can serve as donors of electrons
for the Ti 3\emph{d} orbits in STO, they seem to be different. For example, there
is no carrier freeze-out effect for NSTO even in the most lightly doped
samples because the large dielectric constant of STO makes the activation
energy of a hydrogenic-type donor quite small compared to the thermal energy
at RT [14]. However, in reduced STO, the carrier freeze-out [30,31] can happen
for the electrons originating from doubly charged donor centers - oxygen vacancies,
\emph{i.e.}, the carrier density decreases dramatically with decreasing temperature and
finally the whole system turns to be insulating. The donor level of oxygen vacancies
is large, up to 80 meV separated from the conduction band of STO [14]. This reveals
that the Ti 3\emph{d} electrons originating from oxygen vacancies are naturally more inactive and localized.

As recently reported, localized Ti 3\emph{d} electrons in STO can serve as magnetic centers to
account for Kondo scattering [25] and ferromagnetism [32] in the LaAlO$_{3}$/SrTiO$_{3}$ system. Moreover, oxygen vacancies in the TiO$_{2}$ layer of STO
can enhance the tendency for ferromagnetism considerably [33]. It seems that sufficient
oxygen vacancies as well as carrier density are essential to produce this kind of
indirect ferromagnetic exchange interaction. A possible mechanism for the ferromagnetism closely related to carrier density is the RKKY interaction [34].  Theoretically, the ferromagnetic RKKY interaction under the Weiss mean-field treatment can be strengthened by the increase of carrier density [35,36]. In the free-electron approximation, the RKKY exchange integral oscillates with the distance of magnetic ions at a period of $\lambda$\emph{$_{F}$}/2 [37,38], where $\lambda$\emph{$_{F}$} is the Fermi wavelength. The length scale $\lambda$\emph{$_{F}$}/2 for 0.5 wt\% NSTO single crystals is 1.93 nm [20]. Considering a 0.5 wt\% NSTO single crystal vacuum-annealed at 950 $^{\circ}$C for 1 h, the diffusion length of oxygen vacancies is $\sim$6 $\mu$m and the distance among oxygen vacancies is estimated to be 2.53 nm [20], which is comparable with the RKKY interaction length so that oxygen vacancies could be able to interact with each other by the mediation of a large number of free electrons from Nb doping. However, the indirect exchange coupling between impurity moments in semiconductors via the RKKY interaction is typically weak and yields ferromagnetism with a Curie temperature below room temperature [39,40]. The high Curie temperature in this system is still puzzling to be understood by the RKKY model. The normalized ferromagnetic moment for each Ti$^{3+}$ center originating from oxygen vacancies is estimated to be $\sim$0.05 $\mu_{B}$ [20],which is of the same order with the ferromagnetic moment of localized Ti 3\emph{d} electron at the LaAlO$_{3}$/SrTiO$_{3}$ interface [41]. In addition, no signature of ferromagnetism was observed in the scanning SQUID study of NSTO films in Ref. 41, which is likely because the NSTO samples were annealed at 900 $^{\circ}$C in the oxygen atmosphere to remove oxygen vacancies [42].

In summary, we studied the electrical and magnetic properties of NSTO single crystals. Reversible RTFM was observed
in highly-doped ($\geq$ 0.5 wt\%) NSTO single crystals and found to be induced by oxygen vacancies and closely related
to free carriers. Ferromagnetism moments were found to mostly exist in the surface region of NSTO single crystals. Hysteretic MR as well as the important role of carrier density indicates the intrinsic signature of the ferromagnetism. The use of this kind of substrate to search for novel ferromagnetism in oxide thin films should be exercised with care due to the existence of ferromagnetism up to RT. Even though the ferromagnetic signal observed in the NSTO single crystals is weak, it is strong enough to interfere with magnetic signals of thin films grown on it.

We would like to acknowledge Prof. J.M.D. Coey for enlightening discussions. We thank the National Research Foundation (NRF) Singapore under the Competitive Research Program (CRP) "Tailoring Oxide Electronics by Atomic Control" (Grant No. NRF2008NRF-CRP002-024), the National University of Singapore (NUS) for a cross-faculty grant and FRC (ARF Grant No. R-144-000-278-112) for financial support.


\end{document}